\documentclass[conference]{IEEEtran}
\usepackage{latexsym}
\usepackage{amsfonts}
\usepackage{amssymb}
\usepackage{amsbsy}
\usepackage{amsmath}
\usepackage[dvips]{epsfig}

\title{Random Basketball Routing \\for ZigBee based Sensor Networks}

\author{\IEEEauthorblockN{Dong Min Kim, Young Ju Hwang, Seong-Lyun Kim}
\IEEEauthorblockA{School of Electrical and Electronic Engineering\\
Yonsei University\\
134 Sinchon-Dong, Seodaemun-Gu, Seoul 120-749, Korea\\
Email: \{ verve, promarin, slkim \}@yonsei.ac.kr} \and
\IEEEauthorblockN{Gwang-Ja Jin, Bong-Soo Kim}
\IEEEauthorblockA{RFID/USN Research Group\\
ETRI\\
Yusung P.O Box 106, Daejeon, 305-600, Korea\\
Email: \{ gjjin, bskim \}@etri.re.kr} }

\begin{document}
\maketitle

\begin{abstract}
Random basketball routing (BR) \cite {Hwang} is a simple protocol
that integrates MAC and multihop routing in a cross-layer optimized
manner. Due to its lightness and performance, BR would be quite
suitable for sensor networks, where communication nodes are usually
simple devices. In this paper, we describe how we implemented BR in
a ZigBee-based (IEEE 802.15.4) sensor network. In \cite{Hwang}, it
is verified that BR takes advantages of dynamic environments (in
particular, node mobility), however, here we focus on how BR works
under static situations. For implementation purposes, we add some
features such as destination RSSI measuring and loop-free procedure,
to the original BR. With implemented testbed, we compare the
performance of BR with that of the simplified AODV with CSMA/CA. The
result is that BR has merits in terms of number of hops to traverse
the network. Considering the simple structure of BR and its possible
energy-efficiency, we can conclude that BR can be a good candidate
for sensor networks both under dynamic- and static environments.
\end{abstract}

\begin{keywords}
random basketball routing, relay probability, wireless sensor
networks, IEEE 802.15.4.
\end{keywords}

\section{Introduction}
Wireless sensor networks are composed of a number of cooperative
sensor nodes that are usually powered by batteries, equipped with
small memory components, and operated by poor processing units.
Therefore, energy, memory, and computation efficiency must be
considered for designing communication protocols in a sensor
network. In particular, the routing protocol should be simple and
light. Self-configurability is also required, because any node in
the network can disappear or break down. If a fixed path contains
broken nodes, transmission is failed and energy efficiency may be
severely deteriorated by retransmission through the broken path.

Random basketball routing (BR) is a {\it simple} per-hop-based
multihop routing that incorporates node mobility into the routing
design \cite{Hwang}. In BR, a node (possibly moving) may receive and
forward the same packet multiple times. BR is fully
self-configurable in the sense that the next forwarder (relay node)
is determined adaptively (opportunistically), without knowledge of
the entire network topology. BR {\it integrates} the media access
control (MAC) and routing in a {\it cross-layer optimized} manner.
This intrinsic feature makes BR attractive to sensor networks, where
the sensor is a simple device in most cases.

In BR, there is a key parameter called {\it relay probability}, $p$.
For a given time slot, a node having data either transmits its own
packets with probability $1-p$ (transmitting) or relays for the
other nodes with probability $p$ (listening). When transmitting, a
node sends its packets either to a relay node or directly to the
destination, whichever is appropriate (for example, closest). By
simply controlling the relay probability, we can handle not only
routing but also MAC of the network. When $p=0$, an extreme case,
there is no relay and the routing is reduced to the single-hop
transmission, where every node simultaneously transmits. As the
probability $p$ increases, there are more relay nodes (i.e., less
transmitting nodes) around a transmit node, reducing the average
transmission distance and the delay from retransmissions. However,
the transmission probability $1-p$ of a node also decreases and the
opportunity for the transmission itself becomes smaller. In the
other extreme case, $p=1$, no node is transmitting. Therefore, the
optimal relay probability exists, under which the maximum network
throughput can be obtained.

In \cite{Hwang}, we analyzed the optimal relay probability as a
function of the link data rate and the node density over the service
area. According to our simulations and theoretic investigations,
with BR, the average end-to-end throughput has been improved as the
average speed of nodes increases (as was also noted by \cite{Tse}).
The higher the node density is, the more the gain from node mobility
is.

The purpose of this paper is how we implement BR in a real
ZigBee-based (IEEE 802.15.4) sensor network. Our main emphasis is
how BR works under static situations and we test its performance
using our testbed network. In this paper, we describe how we design
signaling processes, packet structure/type, and other implementing
details. Our experiments are encouraging in that BR outperforms the
conventional AODV (ad hoc on-demand distance vector) routing with
CSMA/CA (carrier sensing multiple access/collision avoidance) MAC.

\section{Signaling, Frame Structure and Loop-free Procedure}

\subsection{Signaling Process}

\begin{figure}[t]
\centerline{\epsfig{figure=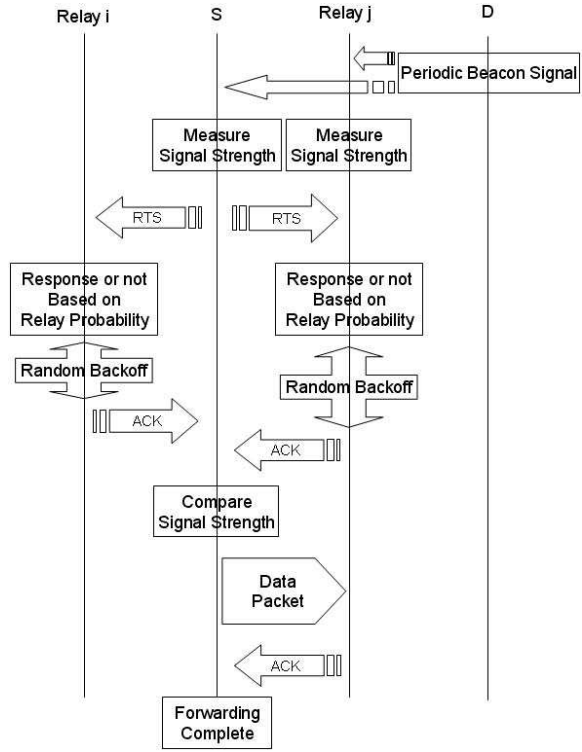,width=3.0in}} \caption{Signaling
for choosing a relay} \label{F:flow}
\end{figure}

\begin{figure}[t]
\centerline{\epsfig{figure=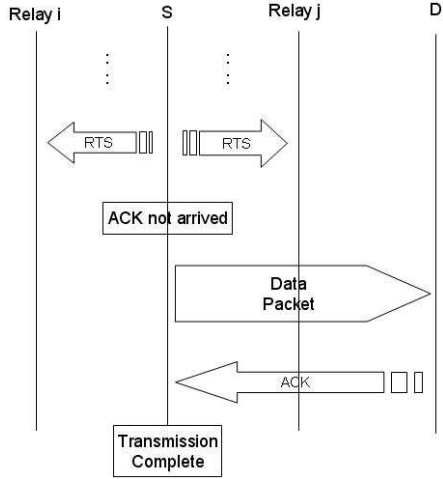,width=2.3in}} \caption{Direct
transmission scenario} \label{F:flow2}
\end{figure}

\begin{figure}[t]
\centerline{\epsfig{figure=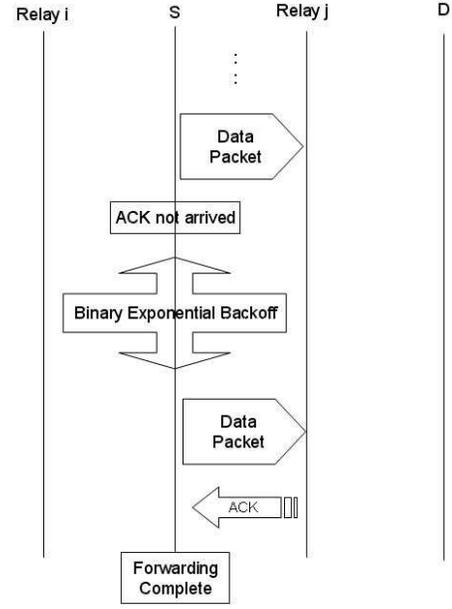,width=2.3in}} \caption{Binary
exponential backoff} \label{F:flow3}
\end{figure}

Figure \ref{F:flow} shows the signaling process of BR for choosing a
relay. Destination node $D$ periodically broadcasts a beacon signal
so that the strength of the signal is measured and stored by each
receiving node. When transmit node $S$ wants to transmit some
packets, it sends out Request-to-Send (RTS) to its neighboring relay
nodes within the radio range. In the RTS frame header, there should
be the identification (ID) of node $S$, the owner of the RTS. Each
neighboring relay node (determined by the relay probability $p$),
after receiving RTS, waits short random time slots to avoid
collision, and then sends out acknowledgement (ACK). In the ACK
packet header, there should be at least two information: one is the
measured signal strength of the periodic beacon signal from the
destination and the other is the ID of the relay node. After
receiving ACK packets from neighboring relay nodes, node $S$
compares them and chooses a relay node (pass) that reports the
strongest received beacon signal. If node $S$ does not receive any
ACK, it sends its packets directly to the destination (shoot) as
shown in Figure $\ref{F:flow2}$, usually not expecting any success
of delivering the packet to its destination.

When node $S$ transmits to a relay node or the destination, a packet
can be transmitted successfully on the condition that the received
SIR at a receiving node is not less than the target SIR $\gamma$. If
the transmission fails, the network adopts the Binary Exponential
Backoff (BEB) \cite{Kwak} for collision resolution (Figure
\ref{F:flow3}). In BR, such retransmission occurs independently of
the probability $1-p$, after the random backoff slots.

\subsection{Packet Structure}

\begin{figure} [t]
\centerline{\epsfig{figure=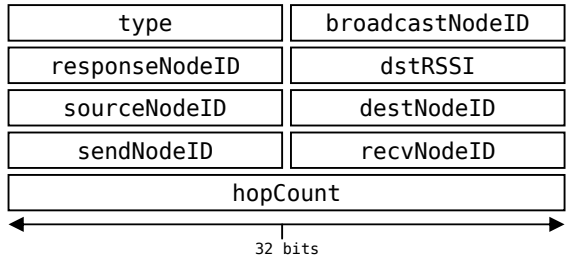,width=3.0in}}
\caption{Basketball routing packet structure}
\label{F:packet_structure}
\end{figure}

Figure \ref{F:packet_structure} shows our design of a BR packet,
where we assigned 16 bits to each field of the packet, except the
\texttt{hopCount} field (32 bits). Tables \ref{T:packet_structure}
and \ref{T:field_description} contain message types of a BR packet
and brief description of each packet field, respectively.

When sensors are powered on, they listen to a channel to receive
packets. A \texttt{TYPE\_DSTBCAST} message is periodically
broadcasted by a destination so that the destination informs other
nodes of its own location. After receiving \texttt{TYPE\_DSTBCAST},
each node measures $RSSI$ and stores it into \texttt{dstRSSI}. A
source node broadcasts a \texttt{TYPE\_SRCBCAST} message to choose
the next forwarder. The neighboring nodes, after receiving
\texttt{TYPE\_SRCBCAST} decide whether to send out
\texttt{TYPE\_RESPONSE} or not, depending on their relay
probabilities. Then, the source node compares its own
\texttt{dstRSSI} and those in received \texttt{TYPE\_RESPONSE}. If
the source's \texttt{dstRSSI} is the largest, the source node
transmits \texttt{TYPE\_ROUTING} directly to the destination.
Otherwise, the source node sends \texttt{TYPE\_ROUTING} to the relay
node which has the largest value of \texttt{dstRSSI}. The node which
receives \texttt{TYPE\_ROUTING} sends a \texttt{TYPE\_ACK} packet to
the source node, and checks if \texttt{destNodeID} is same as its
local address. If then, the routing is completed, otherwise the
relay node who has received \texttt{TYPE\_ROUTING} broadcasts
\texttt{TYPE\_SRCBCAST} and the above procedure is repeated.

\begin{table} [t]
\centerline{
\begin{tabular}{|l|l|l|} \hline
    \em Message Type &\em Description &\em Elements\\ \hline
    \texttt{TYPE\_SRCBCAST} & RTS message & \texttt{broadcastNodeID}\\
    \texttt{TYPE\_DSTBCAST} & Location informing message & \texttt{broadcastNodeID}\\
    \texttt{TYPE\_RESPONSE} & ACK to RTS message & \texttt{broadcastNodeID}\\
                            &                    & \texttt{responseNodeID}\\
                            &                    & \texttt{dstRSSI}\\
    \texttt{TYPE\_ROUTING} & Data message & \texttt{sourceNodeID}\\
                           &              & \texttt{destNodeID}\\
                           &              & \texttt{sendNodeID}\\
                           &              & \texttt{recvNodeID}\\
                           &              & \texttt{hopCount}\\
    \texttt{TYPE\_ACK} & ACK to data message & \texttt{responseNodeID}\\ \hline
\end{tabular}}
\setlength\abovecaptionskip{.25ex plus .125ex minus .125ex}
\setlength\belowcaptionskip{.25ex plus .125ex minus .125ex}
\caption{Message type of a basketball routing packet}
\label{T:packet_structure}
\end{table}

\begin{table}[t]
\centerline{
\begin{tabular}{|l|l|} \hline
    \em Message Field &\em Description \\ \hline
    \texttt{type} & Message type \\
    \texttt{broadcastNodeID} & Broadcasting node ID \\
    \texttt{responseNodeID} & Responding node ID\\
    \texttt{dstRSSI} & Destination message RSSI\\
    \texttt{sourceNodeID} & Message generator ID \\
    \texttt{destNodeID} &  Message receiver ID \\
    \texttt{sendNodeID} & Current forwarder ID \\
    \texttt{recvNodeID} & Next forwarder ID\\
    \texttt{hopCount} & Number of forwardings (relays)\\ \hline
\end{tabular}}
\setlength\abovecaptionskip{.25ex plus .125ex minus .125ex}
\setlength\belowcaptionskip{.25ex plus .125ex minus .125ex}
\caption{Basketball routing packet field description}
\label{T:field_description}
\end{table}


%

\subsection{Loop-free Procedure}

\begin{figure}[tb]
\centerline{\epsfig{figure=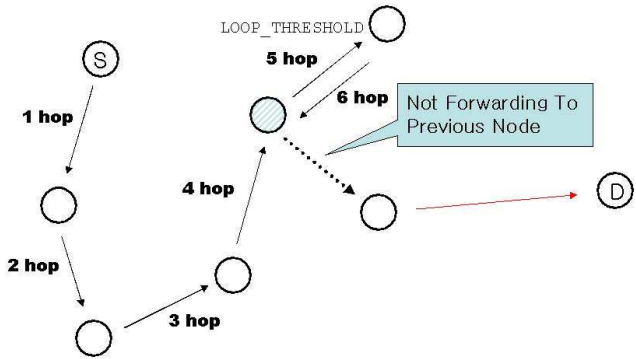,width=3.3in}}
\caption{Message-looping resolution in static networks}
\label{F:loop_resolution}
\end{figure}

With all the merits of BR, there are also some drawbacks from using
it. One problem is that there might be some loops on the routing
paths created by BR and traffic would circulate within a loop. This
can occur in the static environment, because BR allows a node to
relay the same packet more than once. To avoid such loops, if
\texttt{hopCount} exceeded \texttt{LOOP\_THRESHOLD} at a node, a
simple loop-free mechanism is applied; The node does not forward the
packet to a specific node that previously forwarded the same packet
to the current node. The procedure specification is presented in
Figure \ref{F:loop_resolution}.

\section{IEEE 802.15.4 ZigBee Testbed Results}


\begin{table}
\centerline{
\begin{tabular}{|l|l|} \hline
    Parameter & Value \\ \hline
    Number of nodes & 10 \\
    \texttt{RESPONSE\_WAIT\_TIME} & 5 sec \\
    \texttt{ACK\_WAIT\_TIME} & 2 sec \\
    \texttt{BCAST\_TIME} & 10 sec \\
    Relay probability & 0.73 \\
    \texttt{LOOP\_THRESHOLD} & 10 hop \\
    Transmission range & 30 M \\
    Experiment space size & 11.25 M $\times$ 4.05 M \\ \hline
\end{tabular}}
\caption{Experiment parameters} \label{T:experiment_parameter}
\end{table}

\begin{figure}[b]
\centerline{\epsfig{figure=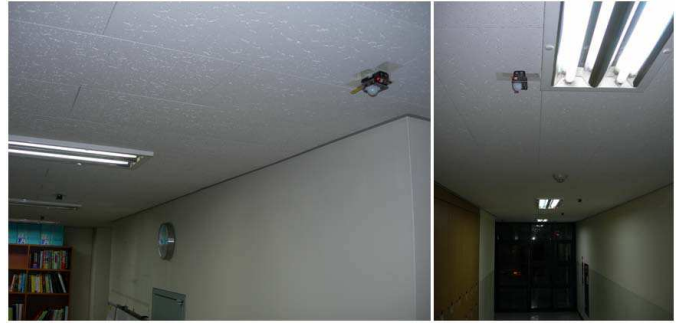,width=3.5in}}
\caption{ZigBee testbed with motion sensors} \label{F:testbed}
\end{figure}

\begin{figure}[!b]
\centerline{\epsfig{figure=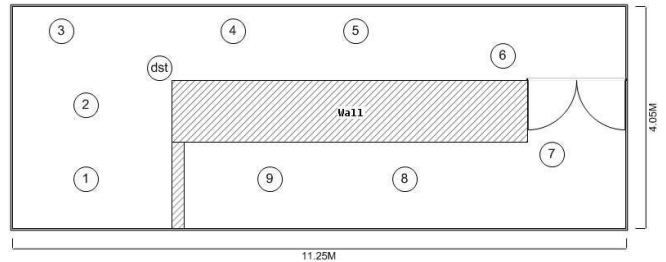,width=3.4in}}
\caption{Testbed topology} \label{F:testbed_topology}
\end{figure}

We have implemented BR on a ZigBee-based (IEEE 802.15.4) testbed.
All hard- and softwares in our experiments are based on $ZigbeX$
bundle manufactured by Hanback Electronics. Each sensor node runs
TinyOS and is equipped with 8-bit $RISC$ ATmega128 and CC2420 chip
as a microcontroller and a radio transceiver, respectively. Sensors
operate on 2.4GHz and communicate at the maximum capacity of 250
kbps within the transmission range.

\subsection{Motion-Detection Sensor Testbed}

Figures  \ref{F:testbed} and \ref{F:testbed_topology} show the network topology of our
indoor testbed. Parameters for experiments are specified in Table \ref{T:experiment_parameter}.
The shaded region represents walls and there is a door on the right side. In our experiments, each node (1-9) senses
the motion of objects and reports the sensed data to the destination
(dst). The destination node is connected to a laptop computer and the
packet arrival from each sensor node is displayed on the computer.
Nodes are distributed so that they can cover all over the network
area with the minimum number of sensors. In this case, we used the optimal relay probability $p^*=0.73$
derived in \cite {Hwang} for the node density $\lambda=0.2$ (number of nodes per unit square meter).
When a node having data is activated with probability $1-p$, BR is executed to transmit the data to the
destination or a relay node, whichever is closest.

\begin{figure}[t]
\centerline{\epsfig{figure=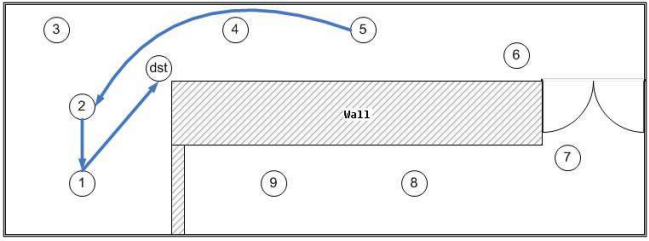,width=3.4in}}
\caption{Route from sensor $5$ to the destination}
\label{F:experiment1}
\end{figure}

\begin{figure}[!t]
\centerline{\epsfig{figure=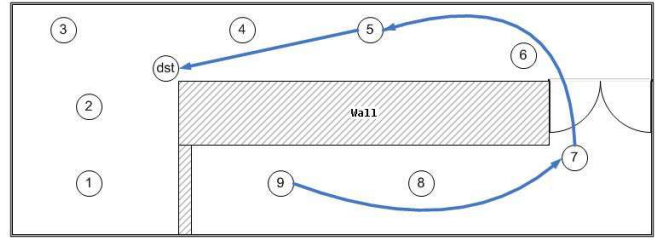,width=3.4in}}
\caption{Route from sensor $9$ to the destination}
\label{F:experiment3}
\end{figure}

Figures \ref{F:experiment1} and \ref{F:experiment3} show the routes
of data sensed at two different nodes. In Figure
\ref{F:experiment1}, the routing succeeded in three hops. With the
other routing protocols using the shortest path as a metric, just
two hops (5-4-dst) would be needed. In BR, however, a node can
receive data from the other nodes only when it acts as a relay with
probability $p$. Therefore, if all intermediate nodes between a
source-destination pair transmit their own data with probability
$1-p$, the next hop (e.g., node 4 in Figure \ref{F:experiment1})
does not respond to the source node (node 5). This might increase
the delay, yet the route taking a roundabout way does not always
have a negative effect on the throughput, as shown in Figure
\ref{F:experiment3}.

In Figure \ref{F:experiment3}, transmission is completed in three
hops, going round to the door. This takes more number of hops
traveling a long distance than the shortest path transmission which
needs only one hop (from node 9 directly to the destination).
However, with the shortest path, delay from retransmissions might be
considerably large because of packet failures, since the destination
is shadowed by the wall. In this case, going round via some nodes by
BR would be better in the delay performance. Like in this case, if
there are obstacles between nodes or a static route is impossible
due to node failures (e.g., battery discharge, sensor breakdown),
then a certain per-hop-based routing is required.

\subsection{BR versus AODV+CSMA/CA on a Tandem Network}

\begin{figure}[tb]
\centerline{\epsfig{figure=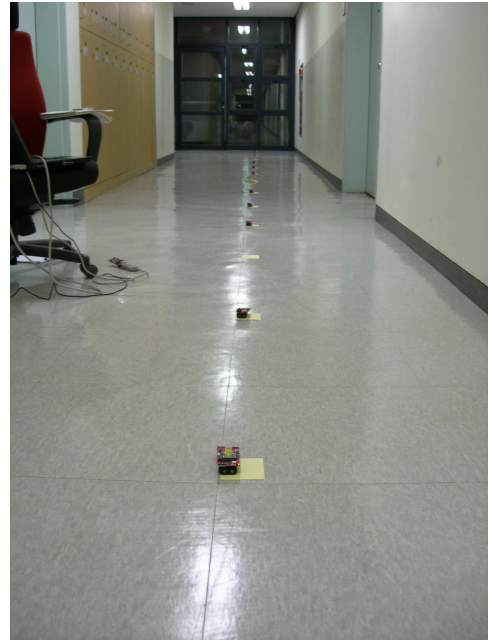,width=2.5in}}
\caption{Tandem network testbed}
\label{F:br_csma_tandem}
\end{figure}

\begin{figure}[tb]
\centerline{\epsfig{figure=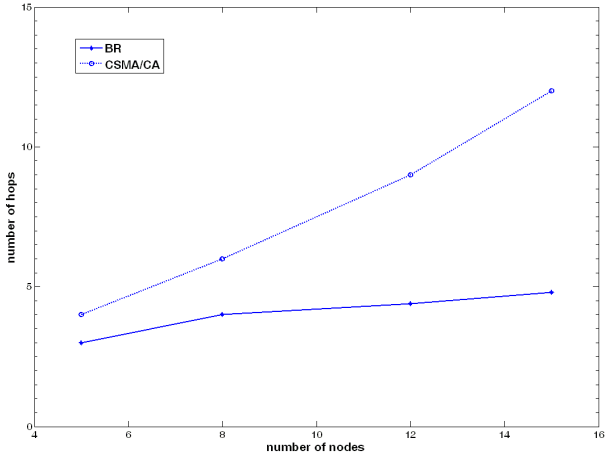,width=3.8in}}
\caption{Number of hops required to route a packet from
source to destination} \label{F:br_csma_hop}
\end{figure}

\begin{figure}[tb]
\centerline{\epsfig{figure=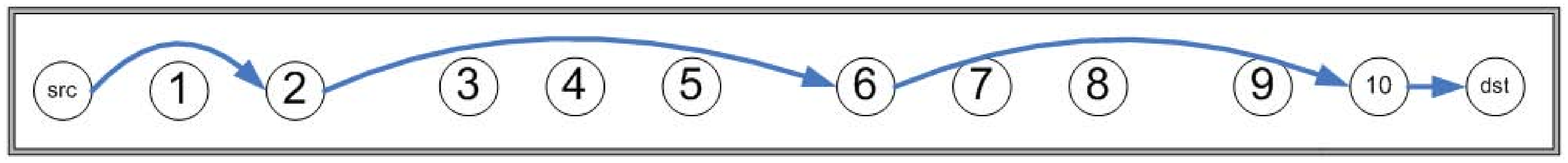,width=3.5in}}
\caption{Route trace in basketball routing} \label{F:br_tandem}
\end{figure}

\begin{figure}[!t]
\centerline{\epsfig{figure=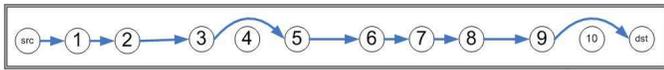,width=3.5in}}
\caption{Route trace in simplified AODV with CSMA/CA}
\label{F:csma_tandem}
\end{figure}

\begin{figure}[t]
\centerline{\epsfig{figure=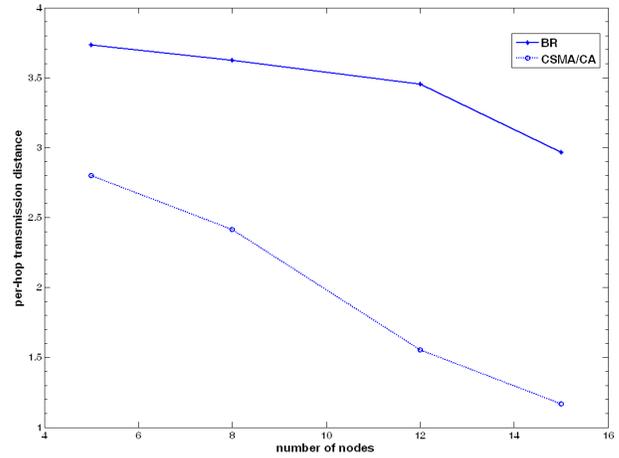,width=3.8in}}
\caption{Per-hop transmission distance as a function of number of nodes}
\label{F:br_csma_distance}
\end{figure}

BR {\it integrates} MAC- and network layer into one, making itself
light and simple. In this subsection, we compare the performance of
BR with that of a simplified AODV+CSMA/CA. We apply a simplified
AODV routing, where the node with the largest RSSI is chosen as the
next forwarder. To determine the effectiveness of BR, we perform
experiments on a tandem wireless network as shown in Figure
\ref{F:br_csma_tandem}. The network contains 5 to 15 nodes
positioned linearly on a $2.05M \times 14M$ floor. The source and
the destination are located at both ends, where the relay nodes are
located between them. The transmission range of a node is set to 6M
and the optimal relay probability $p^*=0.83$. The other parameters
are the same as those in Table \ref{T:experiment_parameter}.

Figure \ref{F:br_csma_hop} compares the average number of hops
required to route a packet from the source to the destination. The
results show that BR performs better than the simplified
AODV+CSMA/CA. Moreover, the larger the network is, the greater the
performance improvement is. In case of the network with 12 nodes
(Figures \ref{F:br_tandem} and \ref{F:csma_tandem}), almost every
node relays packets when the simplified AODV is executed. In
contrast, with BR, it is possible to skip some nearby nodes due to
its opportunistic nature. Similar results are also shown in
\cite{Biswas}. Figure \ref{F:br_csma_distance} shows that as the
number of nodes increases, per-hop transmission distance decreases
in both cases. However, BR utilizes the longer transmission
distance.

\section{Conclusions}
In this paper, we described how we implemented our random basketball
routing (BR) in a ZigBee-based sensor network. Even if BR gets
benefits of dynamic environments, we tested how it works under
static situations, by adding some features like destination RSSI
measuring and loop-free procedure, to the original BR. With BR, at a
given time slot, each source node having data can transmit or
receive packets according to the relay probability. A transmit node
sends its packet to a relay node in the relaying region or directly
to the destination. When transmission fails with a certain level of
collision probability, the Binary Exponential Backoff (BEB) is
executed for collision resolution. BR can be easily applicable in
real sensor networks. Especially, its self-organizing feature can
solve problems like node disappearance or node disorder. Current
work on BR implementation includes comparison with other routing
protocols, measuring actual lifetime of the network both under
dynamic- and static environments.

\section*{Acknowledgment}

This work was supported jointly by Electronics and
Telecommunications Research Institute (ETRI), and the center for
Broadband OFDM Mobile Access (BrOMA) at POSTECH through the ITRC
program of the Korean MIC, supervised by IITA.
(IITA-2006-C1090-0603-0037)

%
%
%
%
%


\begin{thebibliography}{99}

\bibitem{Hwang}
Y. J. Hwang and S.-L. Kim,
\newblock ``The capacity of random wireless networks,"
\newblock {\it submitted for publication},
\newblock 2007.

\bibitem{Tse}
M. Grossglauser and D. N. C. Tse
\newblock ``Mobility increases the capacity of ad hoc wireless networks,"
\newblock {\em IEEE/ACM Trans. Networking},
\newblock Vol. 10,
\newblock No. 4,
\newblock pp. 477-486,
\newblock 2002.

\bibitem{Kwak}
B.-J. Kwak, N.-O. Song and L. E. Miller,
\newblock ``Performance analysis of exponential backoff,"
\newblock {\it IEEE/ACM Trans. Networking},
\newblock Vol. 13,
\newblock No. 2,
\newblock  pp. 343-355,
\newblock 2005.

\bibitem{Biswas}
S. Biswas and R. Morris,
\newblock ``ExOR: Opportunistic multi-hop routing for wireless networks,"
\newblock {\it ACM SIGCOMM},
\newblock 2005.

\end{thebibliography}
\end{document}